\documentstyle[prl,twocolumn,aps]{revtex}
\begin{document}
\title{All consistent interactions for exterior form gauge fields}

\author{Marc Henneaux\,$^{a,b}$\ and  Bernard Knaepen\,$^{a}$}

\address{$^a$ Facult\'e des Sciences, Universit\'e Libre de Bruxelles,
Campus Plaine C.P. 231, B--1050 Bruxelles, Belgium.\\
$^b$ Centro de Estudios Cient\'\i ficos de Santiago, Casilla 16443,
Santiago 9, Chile.\\[.3cm]
\begin{minipage}{14.1cm}\rm\quad We give the complete list of all
first-order consistent interaction vertices for a set of exterior
form gauge fields of form degree $>1$, described in the free limit by
the standard Maxwell-like action.  A  special attention is paid to
the interactions that deform the gauge  transformations.  These are
shown to be necessarily of the Noether form ``conserved antisymmetric
tensor" times ``$p$-form potential" and exist only in particular
spacetime dimensions.  Conditions for consistency to all orders in the
coupling constant are given.  For illustrative purposes,  the
analysis is carried out explicitly for a system of forms with  two
different degrees $p$ and $q$ ($1<p<q<n$).
\end{minipage} }

\maketitle

\pacs{11.17.+y; 11.10.Kk; 02.40.+m}

\narrowtext

It has been known for some time that electromagnetism has a
generalization to
$n$-index antisymmetric tensor potentials, $n=1$ being 
electromagnetism, and that these potentials couple naturally to $n-1$
dimensional extended objects.  These potentials and extended objects
arise throughout supergravity and string theory.  There are many
partial results known about their interactions, but to our knowledge,
there is no systematic treatment of this subject.  In this Rapid
Communication, we provide the complete list of all  interactions that
are consistent to first order in the coupling constant, for an
arbitrary system of non-chiral forms of degrees $>1$, described in
the free limit by the standard Maxwell-like action
\begin{equation} I[A^{(k)}] = - \sum_k  \frac{1}{2(p_k + 1)!} 
\int F^{(k)} \wedge
\overline{F}^{(k)}.
\label{FreeAction}
\end{equation} Here, $F^{(k)}$ is the ``curvature" of $A^{(k)}$,
$F^{(k)} = d A^{(k)}$, and $\overline{F}^{(k)}$ is the dual of
$F^{(k)}$. Furthermore, $p_k$ is the form-degree of the exterior form
$A^{(k)}$ ($2 \leq p_k \leq n-1$).  We also indicate how the analysis
can be pursued to all orders in the coupling constant and explicitly
treat the case when $p_k$ takes only two values
$p, q$, with $2 \leq p < q \leq n-1$ (i.e., the forms in the theory
are either $p$-forms or $q$-forms).  The form degree is taken to be
greater than one in order to exhibit the features peculiar to
exterior forms of higher degree.  However, as mentioned at the end of
the paper, the interactions given for $r$-forms with $r>1$ remain
valid for $r=1$, but simply fail then to exhaust all consistent
interactions (there are additional ones).

The action (\ref{FreeAction}) is invariant under the abelian gauge
transformations
\begin{equation}
\delta_\Lambda A^{(k)} = d \Lambda^{(k)}.
\label{FreeTransf}
\end{equation} and the equations of motion of the free theory are
\begin{equation} d\overline{F}^{(k)} \approx 0.
\label{FreeEOM}
\end{equation} It is well known that an interaction vertex $V=\int
\mu$, where
$\mu = a d^nx$ is a $n$-form built out of the field components and
their  derivatives, is consistent to first order if and only if its
gauge variation under (\ref{FreeTransf}) vanishes up to a surface
term when the equations of motion (\ref{FreeEOM}) hold.  This can be
seen directly by expressing that the sum $I + g V$ is gauge invariant
up to (and including) order
$g$ under gauge transformations that differ from (\ref{FreeTransf})
by terms of at least order $g$. Thus, the determination of all
consistent interactions is equivalent, to first order in the coupling
constant, to the determination of all the ``observables" given by the
spacetime integral of a local $n$-form.  

Consistent interactions of a given gauge theory may be classified into
three categories: (i) those that do not modify the gauge
transformations; (ii) those that modify the gauge transformations
without changing their algebra; and (iii) those that modify both the
gauge transformations and their algebra.  For the first type, the
gauge variation $\delta_\Lambda V$ of the vertex $V$ vanishes (up to
a surface term) off-shell and not just  on-shell.   For the second
and third types, $\delta_\Lambda V$ vanishes only on-shell,
\begin{equation}
\delta_\Lambda V = \sum_k \int b^{(k)}_{\mu_1 \dots \mu_{p_k}}
\frac{\delta I}{\delta A^{(k)}_{\mu_1 \dots \mu_{p_k}}} d^n x
\label{VariationV}
\end{equation} with $b^{(k)}_{\mu_1 \dots \mu_{p_k}} \not= 0$. The
modification of the gauge transformations is given, to first order in
the coupling constant $g$, by 
\begin{equation}
\delta^{NEW}_\Lambda A^{(k)}_{\mu_1 \dots \mu_{p_k}} =  (d
\Lambda)_{\mu_1 \dots \mu_{p_k}} - g b_{\mu_1 \dots \mu_{p_k}}
\label{GaugeTransf}
\end{equation} since then, the gauge variation $\delta_\Lambda^{NEW}
(I + g V)$ vanishes to order $g^2$.  If $b$ is gauge invariant, the
second variation $\delta_{\Lambda_1} ^{NEW} \delta_{\Lambda_2}^{NEW} 
A^{(k)}$ is of order $g^2$ and the interaction does not modify the
gauge algebra to order $g$ \cite{comment1}.

Interactions of each type exist for a set of free vector fields 
$A^a_\mu$.  The interactions that do not deform the gauge
transformations are given by the functions of the curvatures
$F^a_{\mu \nu} \equiv
\partial_\mu A ^a_\nu - \partial_\nu A ^a_\mu$ and their derivatives 
-- as in the Euler-Heisenberg effective Lagrangian for
electrodynamics --, as well as by the Chern-Simons terms in odd
spacetime dimensions \cite{DJT}. There has been no systematic study
of the interactions of the second type but  an example is given in 3
spacetime dimensions by the Freedman-Townsend vertex \cite{FT}
$\epsilon^{\lambda \mu \nu} \overline{F}^a_\lambda
\overline{F}^b_\mu A^c_\nu f_{abc}$.  This coupling has been studied
recently in \cite{Anco}.  Finally, the Yang-Mills coupling, which
exists in  any number of spacetime dimensions, deforms both the gauge
transformations and the gauge algebra, which is no longer abelian
even on-shell.  

As we shall see, the possible interactions of exterior forms of
higher degree are by contrast much more constrained.  The interaction
vertices that deform the gauge transformations exist only when
special conditions are met on the spacetime dimension and the degrees
of the exterior forms. Furthermore, there is no analog of the
Yang-Mills coupling in the sense that no interaction vertex can
deform (non trivially) the gauge algebra at order $g$. This result
generalizes the analysis of \cite{MH} and has a natural
interpretation if the $p$-forms are viewed as connections for
extended objects \cite{Teitelboim}.

The interaction Lagrangians that do not deform the gauge
transformations are completely exhausted, as in the $1$-form case, by
the polynomials in the field strength components and their
derivatives, as well as by the Chern-Simons terms.   These
interactions are consistent not only to first order, but also to all
orders since $I + g V$ is invariant under (\ref{FreeTransf})  exactly
and not just up to order $g^2$.

The interactions that deform the gauge transformations are described
by  the following theorem, which is our main result.

\vskip .2cm

\noindent {\bf Theorem:} {\it The only first-order consistent
interactions that  deform the gauge transformations are given by the
generalized Noether coupling, i.e., are of the form $\sum_k V^{(k)}$,
\begin{equation} V^{(k)} = \int S^{ (k)\mu_1 \dots \mu_{p_k}}
A^{(k)}_{\mu_1 \dots \mu_{p_k}} d^nx
\label{vertex}
\end{equation} where $S^{ (k)\mu_1 \dots \mu_{p_k}}$ are non-trivial
{\em gauge-invariant} antisymmetric tensors which are conserved modulo
the equations of motion,
\begin{equation}
\partial_{\mu_1} S^{\mu_1 \dots \mu_{p_k}} \approx 0, \;
\delta_\Lambda S^{\mu_1 \dots \mu_{p_k}} = 0.
\label{current}
\end{equation} In form notations,
\begin{equation} V^{(k)} = \int J^{ (k)} \wedge A^{(k)},
\label{vertexBis}
\end{equation} where $J^{ (k)}$ is the non-trivial, gauge-invariant,
conserved ($n-p_k$)-form dual to $S^{ (k)\mu_1 \dots \mu_{p_k}}$,
\begin{equation} d J^{ (k)} \approx 0, \; \delta_\Lambda J^{ (k)} = 0.
\label{currentBis}
\end{equation} }

\vskip .2cm
\noindent {\bf Proof:} That (\ref{vertexBis}), (\ref{currentBis})
define first-order consistent interactions is rather obvious because 
$\delta_\Lambda(J^{ (k)} \wedge A^{(k)}) = J^{ (k)} \wedge 
\delta_\Lambda A^{(k)} = J^{ (k)} \wedge d \Lambda^{(k)}
\approx d(\pm J^{(k)} \wedge \Lambda^{(k)})$, where the $+$ ($-$)
sign occurs when $J^{(k)}$ is a form of even (odd) degree. That these
are the only interactions that deform the gauge symmetries is harder
to prove and is based on the cohomological reformulation of the
observables, which are known to be  in bijective correspondence with
the elements of the BRST cohomology $H^0(s \vert d)$ at ghost number
zero \cite{HPT2,hen91,bbh1}.  One can work out $H^0(s \vert d)$   by
following exactly the pattern developed in \cite{YMcoho} for analysing
the mod $d$ BRST cocycles in the Yang-Mills case, as well as the
results of
\cite{hks} on the characteristic cohomology for exterior gauge fields
described by the action (\ref{FreeAction}). If one does this, one
finds that the antifield-independent part of the antifield-dependent
cocycles (which are the cocycles that deform the gauge symmetry
\cite{BH,GW}) can always be brought to the form given in the theorem.
The details will be reported elsewhere \cite{HK}.  

The major difference between the allowed couplings between exterior
forms of degrees $\geq 2$ and the Yang-Mills cubic coupling, which
may be written as in (\ref{vertex}), but with the {\it non
gauge-invariant} current $f^a_{bc}F^{b\mu \nu} A^c_\nu$, is that the
latter deforms non-trivially the gauge algebra already at order $g$
in the coupling constant, while the former leaves it abelian at that
order.  This is because the vertex (\ref{vertex}) is linear in the
non-gauge invariant form $A$.  Therefore, upon integration by parts,
one finds that $\delta_\Lambda V^{(k)}$ is a combination of the
Euler-Lagrange derivatives of the free Lagrangian with coefficients
that are {\it gauge invariant}. Thus, according to (\ref{VariationV})
and (\ref{GaugeTransf}), the first-order modification of the gauge
transformations is gauge invariant and does not change the abelian
nature of the gauge algebra.  By contrast, the conserved current
entering the Yang-Mills coupling is not gauge invariant.  This
possibility is not present here because the conserved $m$-forms ($m
\leq n-2$) are necessarily gauge invariant up to trivial terms
\cite{hks}.

Since the non-trivial conserved $m$-forms are known to be, for $m \leq
n-2$, the polynomials in the curvature forms $F^{(k)}$ and their
duals $\overline{F}^{(k)}$ \cite{hks}, the theorem provides the most
general interaction vertices that can be added to the free action
consistently to first order
\begin{eqnarray} &&I \rightarrow I + \sum_{(A)} g_{(A)} V_{(A)},
\label{ConsistentInt} \\  V_{(A)} &=& \int \overline{F}^{(k_1)} \dots
\overline{F}^{(k_2)} F^{(l_1)}  \dots  F^{(l_2)}  A^{(t)}.
\label{ConsistentV}
\end{eqnarray} From now on, we shall drop the wedge symbol in
exterior products.  In (\ref{ConsistentV}), $V_{(A)}$ contains  at
least one dual $\overline{F}^{(k)}$ since otherwise it reduces to a
Chern-Simons term and does not deform the gauge symmetry.

A remarkable feature of the first-order vertices  $V_{(A)}$  is that
they are in finite number.  Indeed, one can form only a finite number
of polynomials (\ref{ConsistentV}) of form degree $n$ (we exclude the
rather direct case of forms of degree $n-1$, for which the field
strengths are $n$-forms and the theory has no local degree of freedom,
since the duals are then $0$-forms of which one can take arbitrarily
high powers).  For a given spacetime dimension and a given set of
exterior forms, all the vertices deforming the gauge transformations
can be listed explicitly. There may actually even be no conserved
($n-p_k$)-form that could match the form-degree $p_k$ of $A^{(k)}$ in
(\ref{vertex}) to make an $n$-form, in which case there would be
simply no consistent vertex that would deform the gauge
transformations.  An example is given in
\cite{MH}.

Let us now turn to the consistency of the vertices to higher orders. 
The most expedient way to analyse this question is to rephrase the
problem in terms of the master equation and its deformations
\cite{BH,Stasheff}.  Then, one easily sees that the obstructions to
second (and higher) order consistency lie in the local BRST cohomology
$H^1(s \vert d)$ at ghost number one.  A first-order consistent
deformation is obstructed to second order if its antibracket with
itself, which is BRST-closed, is not BRST-exact
\cite{BH}.  In particular, if the cohomology group 
$H^1(s \vert d)$ vanishes, no first-order consistent interaction can
be obstructed at higher order. This raises the question of computing
$H^1(s \vert d)$. The calculation of $H^1(s \vert d)$ can be
performed and follows exactly that of $H^0(s \vert d)$.   Again, the
detailed calculation will be reported elsewhere
\cite{HK}.

Knowing $H^1(s \vert d)$, one can in principle investigate the higher
order consistency of any given first-order consistent vertices for an
arbitrary system of forms.  The procedure can be rather tedious  in
practice and we shall, for illustrative purposes, consider here only
the explicit case when two form degrees $2\leq p < q \leq n-2$ are
present. We denote the $p$-forms by $A^a$ ($a = 1, \dots, m$) and the
$q$-forms by $B^A$ ($A = 1, \dots, M$), with respective curvatures
$F^a = dA^a$ and $H^A = dB^A$.  These are respectively ($m+1$)- and
($M+1$)-forms, while their duals $\overline{F}^a$ and 
$\overline{H}^A$ are respectively ($n-m-1$)- and ($n-M-1$)-forms.
Taking into account the fact that the interactions (\ref{vertex})
exist only if $n-p$ or $n-q$ can be written as $M_1 + M_2 +M_3 +M_4$,
where $M_1$ is a multiple of $p+1$, $M_2$ is a multiple of $q+1$,
$M_3$ is a multiple of $n-p-1$ and $M_4$ is a multiple of $n-q-1$,
with $M_3 + M_4 \not= 0$ in order to have at least one dual, we have
found that there are only three types of ``basic" first-order
consistent interactions that deform the gauge symmetry:

\vskip .1cm

\noindent (i) Chapline-Manton couplings, which are linear in the duals
\cite{CM},
\begin{eqnarray} V_1 &=& \int f_{aA} \overline{F}^a  B^A, \; \;
(q=p+1),
\label{CM1} \\ V_2 &=& \int f_{Aa_1 \dots a_{k+1}} 
\overline{H}^A F^{a_1}  \dots F^{a_{k}} A^{a_{k+1}}, \nonumber \\ &&
(k(p+1) +p = q+1).
\label{CM2}
\end{eqnarray} Here, $f_{aA}$ and $f_{Aa_1\dots a_{k+1}}$ are
arbitrary constants.  The
$f_{Aa_1\dots a_{k+1}}$ may be assumed to be completely symmetric
(antisymmetric) in the $a$'s if $p$ is odd (even). The
Chapline-Manton coupling (\ref{CM1}) exists only if $q=p+1$; the
Chapline-Manton coupling (\ref{CM2}) exists only if 
$k(p+1) +p = q+1$ for some integer $k$.

\vskip .1cm

\noindent (ii) Freedman-Townsend couplings, which are quadratic in
the duals \cite{FT},
\begin{eqnarray} V_3 &=& \int f^A_{BC} \overline{H}^B  \overline{H}^C
B_A ,
\label{FT1} \\ V_4 &=& \int t^a_{Ab} \overline{H}^A  \overline{F}_a
A^b .
\label{FT2}
\end{eqnarray} Here, $f^A_{BC}$ and $t^a_{Ab}$ are constants that are
arbitrary at first order but will be restricted at second order.
The Freedman-Townsend vertices (\ref{FT1}) and (\ref{FT2}) exist only
if
$q=n-2$.

\vskip .1cm

\noindent (iii) Generalized couplings, which are at least quadratic
in the duals $\overline{H}^A$,
\begin{eqnarray} V_5 = \int k_{A_1 \dots A_l a_1 \dots a_{k+1}}
\overline{H}^{A_1} ...  
\overline{H}^{A_l} F^{a_1} ... F^{a_k}  A^{a_{k+1}}
\label{GenInt}
\end{eqnarray} where $k_{A_1 \dots A_l a_1 \dots a_k a_{k+1}}$ are
arbitrary constants with the obvious symmetries.  These interactions
exist only if there are integers $k$, $l$ (with $l \geq 2$) such that
$l(n-q-1) + k(p+1) +p = n$.

None of the above interactions may be available.  This would occur,
for  instance, for $n=11$, $p=2$, $q=5$, for which there is  thus no
consistent, direct interaction of the 2-forms and the 5-forms that
deforms the gauge symmetries (although these forms may of course
interact through the Chern-Simons terms $F^2 B$ or $FHA$ which do not
deform the gauge symmetries or through the exchange of another field).

The Chapline-Manton first-order coupling (\ref{CM1}) or (\ref{CM2})
defines a  fully consistent interaction that is most easily obtained
by introducing the gauge-invariant field-strengths $F = dA -gfB$ in
the first case or $H= dB - gfF^kA$ in the second case.  The
interacting theory is simply given by the free action in which the
original field strengths are replaced by the gauge-invariant ones. 
This automatically generates the correct $O(g^2)$-terms.  

The Freedman-Townsend vertices define a consistent theory to higher
orders if and only if two conditions are met: (i) the $f^A_{BC}$
fulfill the Jacobi identity and thus define a Lie algebra;  (ii) the
$t^a_{Bc}$ define a representation of that Lie algebra.  These
restrictions arise because $H^1(s \vert d)$ does not vanish and the
antibracket of the cocycles defining the first order interaction is
not zero in cohomology unless (i) and (ii) are fulfilled.  When these
conditions are met, one finds that the fully interacting theory is
given, in first order form for the $B$-fields, by
\begin{eqnarray} I= \int - \frac{a}{2}(2B_A \Phi^A  +
\overline{\beta}^A  \beta_A) - \frac{1}{2 b} \overline{F}'^{a}  F'_a
\label{fullFT}
\end{eqnarray} with $a = (n-1)!(-1)^n/(q+1)!$ and $b = (p+1)!$. In
(\ref{fullFT}), $\beta^A$ is an independent $1$-form  that coincides
on-shell with the dual of
$H^A$ in the free limit, while $\Phi^A$ is its curvature,
$\Phi^A = d \beta^A + (g/2) f^A_{BC} \beta^B \beta^C$, and $F'^a$ is
the covariant exterior derivative of $A^a$,
$F'^a = DA^a \equiv d A^a + g t^a_{Bc} \beta^B \wedge A^c$. The
action (\ref{fullFT}) is invariant under the abelian gauge
transformations
\begin{eqnarray}
\delta B_A = D \Lambda_A - \frac{g}{ab} t^{a}_{Ab} 
\overline{F}'_a \Lambda^b, \, \delta \beta^A = 0,
\, \delta A^a = D \Lambda^a,
\label{GaugeTransfFullFT}
\end{eqnarray} which are still reducible on-shell because $\Phi^A
\approx 0$.  If one eliminates $\beta^A$ by means of its equations of
motion, one gets the second-order form of the Freedman-Townsend
model, which is non polynomial.  If one eliminates instead $B^A$, 
one gets the non-linear sigma-model with a minimal coupling of the
exterior form $A^a$ to the flat connection $g^{-1} dg$.  Note that
the metrics
$\delta_{AB}$, $\delta_{ab}$ with which we have lowered and raised
the internal indices need not be invariant \cite{MH}.

Lastly, we turn to the generalized couplings (\ref{GenInt}).  These
define first-order terms of a fully consistent interaction no matter
how the coefficients $k_{A_1 \dots A_l a_1 \dots a_k a_{k+1}}$ are
chosen.  The corresponding full theory is (in first order form for
$B^A$)
\begin{eqnarray} &I&= \int - \frac{a}{2}(2B_A \Phi^A + 
\overline{\beta}^A  \beta_A) - \frac{1}{2 b} \overline{F}^{a} F_a
\nonumber \\ && + g \int k_{A_1 \dots A_l a_1 \dots a_{k+1}}
\beta^{A_1} \dots \beta^{A_l}  A^{a_1} F^{a_2} \dots F^{a_{k+1}}
\label{FullNew}
\end{eqnarray} with $a = - (-1)^{(n-q-1)(q+1)} (n-q-1)!$ and $b  =
(p+1)!$  and with (reducible) gauge transformations
\begin{eqnarray}
\delta A^a &=& d \Lambda_{a_1}, \; \; \delta \beta^A = 0, 
\; \; \delta B_{A_1} = d \Lambda_{A_1} \nonumber \\ &-& \alpha k_{A_1
\dots a_{k+1}} \beta^{A_2}
\dots \beta^{A_l} \Lambda^{a_1} F^{a_2} \dots F^{a_{k+1}}.
\label{GaugeTransfFullNew}
\end{eqnarray} Here, $\Phi^A = d \beta^A$ and $\alpha = l
(-1)^{(n-q)(l-q)-l-1}/ a$.  Again, upon elimination of the auxiliary
field $\beta^A$, one generates the first-order vertex (\ref{GenInt})
and the corresponding higher-order terms.   This theory does not
appear to have been described explicitly in the previous literature. 
It is dual to a theory with pure Chern-Simons couplings, as it can be
easily seen by eliminating $B^A$ instead of
$\beta^A$.

Note that the algebra of the gauge transformations remains abelian
on-shell to all orders in the coupling models for the above three
models \cite{comment3}.
 
To conclude, we have shown that the gauge symmetries of exterior form
gauge fields have a high degree of rigidity. Interactions that deform
them do exist, but only in special dimensions.  Furthermore, they
never modify the gauge algebra to first order in the coupling
constant. Couplings to $1$-forms can be treated along similar lines. 
One finds additional interactions besides those given by the above
theorem, which are also of the Noether form $j^\mu A_\mu$. However,
the conserved current $j^\mu$ which couples to the
$1$-form need not be gauge-invariant.  There is actually only one non
gauge-invariant current that is available and it leads to the
Yang-Mills cubic vertex, which deforms the gauge algebra to order
$g$.  All other currents $j^\mu$ may be assumed to be gauge-invariant
and thus do not lead to algebra-deforming interactions.  There is in
particular no vertex of the form $\overline{H} B A$ where A are
1-forms and B are
$p$-forms ($p>1$) with  curvature $H$, which excludes charged
$p$-forms (i.e. $p$-forms transforming in some representation of a
Lie algebra minimally coupled to a Yang-Mills potential). An
important difference with the above case is that the gauge-invariant
conserved currents $j^\mu$ are in infinite number because the starting
theory is free and possesses an infinite number of conserved
charges.  There is accordingly an infinite number of first-order
consistent interaction vertices but most of them are of course
inconsistent at higher order.

\vskip .2cm

\noindent {\bf Notes added}

\vskip .2cm

1.  As stated in the conclusion, the interaction vertices of this
paper are still available in the presence of $1$-forms. They just
fail to exhaust then all the possible vertices.   In particular, the
vertices (i), (ii) and (iii) given above for a mixed system of $p$
and $q$-forms are still available when $p=1$; the analysis of their
higher-order consistency is also equally valid.  

In a recent preprint \cite{BrandtDragon},  Brandt and Dragon have
described an interaction between two $1$-forms $A^a$ ($a,b= 1,2$) and
one $2$-form 
$B^1$ in four dimensions that actually fits into the
general Freedman-Townsend structure (\ref{FT1}), (\ref{FT2}) by taking
$f^A_{BC} = 0$ ($A=1$, abelian one-dimensional algebra) and $t^2_{11}
= 1$, other components of $t^a_{1b}=0$.  [In their notations, $A^1 =
W$ and
$A^2 = A$].  It is clear that $[t_1,t_1] =0$, so this $2$ by $2$
matrix defines a representation of the abelian one-dimensional Lie
algebra with $f^A_{BC} = 0$. Therefore, the higher-order consistency
condition are fulfilled and the full interaction is given by Eq.
(\ref{fullFT}).  Elimination of the auxiliary field $\beta$ is easily
checked to reproduce the action of \cite{BrandtDragon}. Thus, this
action precisely falls in what we have called the Freedman-Townsend
category. Note that due to the triangular form of the matrix $t_1$,
one can dualize not only the $2$-form $B^1$ (to get a scalar), but
also the potential $A^2$ (to get another one-form).  If one does so,
one obtains  two $1$-forms and one scalar coupled through a standard
Chern-Simons term. The vertex (\ref{FT2}) specialized to that
peculiar set of fields and to a similar choice of  coefficients was
considered previously  to first order in $g$ by Brandt in the
supergravity context \cite{Brandt}.

Our approach, which covers the general cases of both abelian and
non-abelian Lie algebras as well as arbitrary  representations, and
which also covers higher-degree forms in higher spacetime dimensions,
shows explicitly that the familiar  concepts on which Yang-Mills
theory is based (curvature, covariant derivatives, representations)
are shared by the model (\ref{fullFT}).  It is precisely the
recognition of this similarity that enabled us to construct the
interaction vertex (\ref{fullFT}) to all orders. 

\vskip .2cm

\noindent 2. The basic interaction vertices described above can of
course be combined, or can be combined with Chern-Simons terms.  This
leads, in general, to additional constraints on their coefficients
(which may actually have no non trivial solutions in some cases). 
One example of a non-trivial combination is the description of 
massive vector fields worked out in \cite{FT,Thierry}, which combines
the Freedman-Townsend vertex with a Chern-Simons term \cite{Bresil}. 
Another example is given in \cite{AncoBis}, where both
the Freedman-Townsend vertex and the Yang-Mills vertex are introduced
simultaneously.

\vskip .2cm

B. K. is ``Aspirant du Fonds National de la Recherche Scientifique"
(Belgium).  We thank Friedemann Brandt for useful discussions on the
work \cite{BrandtDragon}.


\begin{references}
\bibitem{comment1} We include in the first category the interaction
vertices that do not fulfill $\delta_\Lambda  V= 0$ off-shell, but
can be made to do so by a field redefinition.  Similarly, we include
in the second category the interaction vertices that deform non
trivially the gauge transformations but deform trivially their
algebra,  in the sense that the gauge algebra can be made abelian by
appropriate redefinition. This will always be understood in the
sequel.
\bibitem{comment3}  With three (or more) different degrees, however,
one may modify the algebra at order $g^2$. For instance, if $A$, $B$
and $C$ are respectively 3-, 4- and 7-forms, the Lagrangian $ \sim
\overline{F} \wedge F +
\overline{H} \wedge H + \overline{G} \wedge G$ is invariant under the
gauge transformations $\delta A = d \epsilon + g \Lambda$,
$\delta B = d \Lambda$ and $\delta C = d \mu + g \epsilon dB - g^2
\Lambda B$, where $\epsilon$, $\Lambda$ and $\mu$ are respectively
2-, 3- and 6-forms.  Here, $F = dA -g B$, $H = dB$ and $G = dC - g A
dB + (1/2) g^2 B^2$.  The commutator of two $\Lambda$-transformations
is a $\mu$-transformation with $\mu = g^2 \Lambda_1 \Lambda_2$. The
fact that the gauge algebra may become non-abelian at order
$g^2$ shows that the connection interpretation, which excludes this
possibility, is not always appropriate.
\end{references}
\end{document}